\documentclass[prl,aps,epsf,twocolumn,showpacs]{revtex4-1}
\usepackage[pdftex]{graphicx}
\usepackage{dcolumn}
\usepackage{bm}
\usepackage{epsfig}
\usepackage{latexsym}
\usepackage{amsmath}
\usepackage{amssymb}
\usepackage{color}
\usepackage{array}

\newcommand{\<}{\langle}
\newcommand{\e}{\varepsilon}
\newcommand{\up}{\uparrow}
\newcommand{\down}{\downarrow}
\renewcommand{\>}{\rangle}
\renewcommand{\(}{\left(}
\renewcommand{\)}{\right)}
\renewcommand{\[}{\left[}
\renewcommand{\]}{\right]}
\renewcommand{\v}[1]{\mathbf{#1}} 
\newcommand{\dslash}{d \hspace{-0.8ex}\rule[1.2ex]{0.8ex}{.1ex}}
\newcommand{\bs}{\boldsymbol}

\begin{document}
\title{Edge-Ferromagnetism from Majorana Flat-Bands: Application to Split Tunneling-Conductance Peaks in the High-Tc Cuprates}
\author{Andrew C. Potter and Patrick A. Lee}
\affiliation{Department of Physics, Massachusetts Institute of Technology, Cambridge, MA 02139, USA}

\begin{abstract} In mean-field descriptions of nodal d-wave superconductors, generic edges exhibit dispersionless Majorana fermion bands at zero-energy.  These states give rise to an extensive ground-state degeneracy, and are protected by time-reversal (TR) symmetry.  We argue that the infinite density of states of these flat-bands make them inherently unstable to interactions, and show that repulsive interactions lead to edge FM which splits the flat bands.  This edge FM offers an explanation for the observation of splitting of zero-bias peaks in edge tunneling in High-Tc cuprate superconductors.  We argue that this mechanism for splitting is more likely than previously proposed scenarios, and describe its experimental consequences.
\end{abstract}
\maketitle

The discovery of topological insulators\cite{TIRMP} has led to a re-examination of the role of symmetries for protecting surface states. Since known topological invariants are defined only in the presence of an energy gap, this effort has focused almost exclusively on gapped insulators and superconductors.  However, well-defined topologically protected surface states can emerge at the boundary of (non-interacting) gapless systems, so long as translational invariance along the boundary is preserved.  For example, in the absence of interactions, clean systems with bulk Dirac nodes, such as graphene\cite{Graphene}, Weyl-semimetals\cite{Weyl1,Weyl2}, and nodal superconductors\cite{Hu,FaDungHai} are all expected to exhibit dispersionless bands that are spatially confined to the edge.  In special cases flat-bands can persist in gapless superconductors even in the presence of disorder\cite{Wong}.  

These flat edge-bands exist in regions of linear size $\Lambda$ in the boundary-Brillouin zone, and terminate at bulk gapless nodes where the edge states delocalize from the boundary.  In superconducting cases, these edge bands are pinned to the Fermi-energy as neutral Majorana fermions, giving rise to large ground-state degeneracy: $D\approx 2^{\(\Lambda/2\pi\)^{d-1}}$, where $d$ is the spatial dimension of the bulk.  This corresponds to an extensive ground-state entropy: $S_0 \sim \(\Lambda L\)^{d-1}$ in violation of the third law of thermodynamics, and also to an infinite density of states at zero-energy.  Therefore even arbitrarily weak interactions act as a singular perturbation, and play an essential role in determining true edge-state structure.

In this paper, we examine the effects of interactions on flat edge-bands, focusing specifically on the case of 2D superconductors with nodal $d_{x^2-y^2}$ pairing symmetry, relevant to the cuprate family of high-temperature superconductors (SC).   The flatness of the edge-states is protected not only by translation symmetry along the edge, but also time-reversal symmetry (TRS).  Therefore, TRS breaking order is required lift the edge degeneracy.  

\begin{figure}[ttt]
\begin{center}
\includegraphics[width = 3.2in]{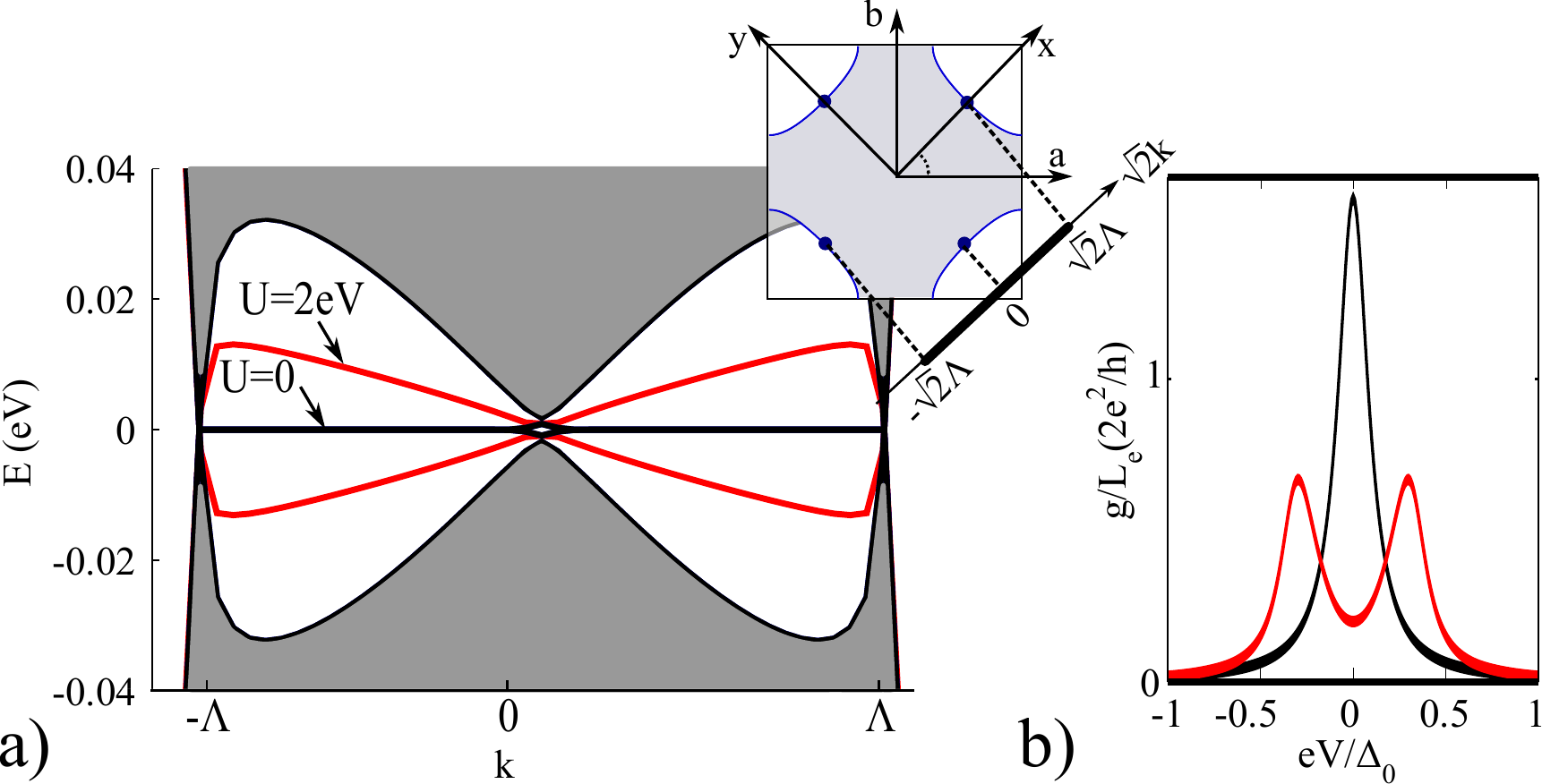}
\end{center}
\vspace{-.2in}
\caption{a) Momentum resolved spectrum for the edge for non-interacting d-wave nodal superconductor for $t=240$meV, $\Delta_0 = 10$meV, and $U=0$ (black-line, non-interacting) and $U=2$eV (red curves, mean-field).  Solid lines indicate edge states and gray, shaded regions represent the bulk continuum.  In the presence of interactions the edge becomes ferromagnetic, splitting the flat band of edge-states. The small splitting at $k=0$ is a finite size artifact. Inset shows the 2D Brillouin zone, rotated coordinate system, and projection onto the edge.  b) Edge tunneling conductance for $\gamma_0=0.1\frac{U}{\pi\xi_0}$, for $U=0$ (black) and $U=2$eV (red). 
}
\vspace{-.2in}
\label{fig:EdgeSpectrum}
\end{figure}

Evidence for such TRS breaking was found in normal-metal/YBCO tunnel junction experiments.  These show a large zero-bias peak at intermediate temperatures\cite{ZBP1,ZBP2,ZBP3}, corresponding to many low-lying states.  The peak subsequently splits into two as the sample is cooled\cite{CovingtonSplitting,Krupke}.  Previously proposed explanations\cite{Matsumoto,Fogelstrom} of this phenomena, were based on the point of view that the edge-scattering is pair-breaking and suppresses the $d_{x^2-y^2}$ parameter near the edge.  In principle, this then allows a different, sub-dominant pairing symmetry to develop near the edge.  In particular, s-wave pairing with a relative $\frac{\pi}{2}$ phase to the bulk order parameter ($d+is$) was suggested to develop near the edge\cite{Matsumoto,Fogelstrom}.  However, the relevance of this scenario to the high-Tc cuprates is questionable, since superconductivity in these materials is widely believed to arise from strong electron-electron repulsion\cite{HighTcRMP,HighTcSpinFluctuations,HighTcDMFT}, which disfavors s-wave pairing\cite{Endnote:Gwave}. 

Viewing the zero-bias peak, instead, from the perspective of topological flat bands naturally suggests a different scenario.  Namely, since the flat edge-bands are spin-degenerate, it is natural to suspect that repulsive interactions will produce ferromagnetism (FM) due to exchange forces, in close analogy to quantum Hall FM in flat Landau-levels\cite{QHFM}.  Here, we confirm this expectation, and show that whereas repulsive interactions disfavor $d+is$ pairing, arbitrarily weak repulsion favors FM at the edge (even if there is no tendency towards FM in the bulk).  We identify tunneling-signatures of the edge-FM, which can distinguish it from previously proposed $d+is$ pairing.   

\vspace{6pt}\noindent\textbf{\textit{Flat Edge Bands - }}
To set the stage, we review the structure of flat edge bands in the absence of interactions, as previously discussed in \cite{FaDungHai}.  We start by considering the mean-field Hamiltonian of a superconductor with $d_{x^2-y^2}$-pairing written as a tight-binding model on the square lattice. In the absence of an edge, the Hamiltonian can be written in momentum space:
\begin{align} \label{eq:Hdwave} 
H_d &= \frac{1}{2}\sum_{k}\Psi_k^\dagger\[\mathcal{H}_0(\v{k})\tau_3+\mathcal{H}_\Delta(\v{k})\tau_1\]\Psi_k
\nonumber \\
\mathcal{H}_0&=-2t\(\cos k_a+\cos k_b\)-\mu
\nonumber\\
\mathcal{H}_\Delta &=-2\Delta_0\(\cos k_a-\cos k_b\)
\end{align}
(see Fig.~\ref{fig:EdgeSpectrum} inset) where $\Psi = \begin{pmatrix} c_{\up,\v{k}} & c_{\down,-\v{k}} & c_{\down,-\v{k}}^\dagger & -c_{\up,\v{k}}^\dagger \end{pmatrix}$.  Furthermore, we ignore the effect of phase fluctuations, which will not be important in what follows, and choose $\Delta_0$ to be uniform and real.  For convenience, we have chosen units of length such that the lattice spacing is unity, and shifted the chemical potential such that $\mu=0$ corresponds to half-filling.

In anticipation of introducing an edge along the $\<11\>$ direction, we re-write the Hamiltonian in terms of momenta along, $k = \frac{k_a+k_b}{2}$, and perpendicular, $k_\perp = \frac{k_a-k_b}{2}$, to the edge, where $k,k_\perp\in[-\frac{\pi}{\sqrt{2}},\frac{\pi}{\sqrt{2}}]$.  In these coordinates we have:$\mathcal{H}_0 = -2t_{k}\cos(k_\perp)-\mu$, and $\mathcal{H}_\Delta = 2\Delta_{k}\sin(k_\perp)$
where $t_k = 2t\cos k$ and $\Delta_k = 2\Delta_0\sin k$.  

Next, we introduce an edge.  Though any direction misaligned from $\<10\>$ will exhibit edge bands, for concreteness we consider an edge along the $\<11\>$ direction.  Since $k_\perp$ is no longer a good quantum number, one must move to a real-space description for the y-direction.  For each value of $k$, the Hamiltonian is formally identical to an effective 1d tight-binding chain with hopping, $t_k$, and p-wave pairing $\Delta_k$.

It is well known \cite{KitaevWire} that such 1d wires with p-wave pairing exhibit zero-energy Majorana end-states so long as: $|2t_k|<|\mu|$ and $\Delta({k})\neq 0$.  In the present case, there will be a Majorana zero-mode for each $|{k}|<\cos^{-1}\frac{\mu}{4t}$, ${k}\neq 0$.  These edge-bands terminate at bulk nodal points: $k = 0,\pm \Lambda$, where $\Lambda = \cos^{-1}\frac{\mu}{4t}$.

For ${k}$ between the bulk nodes, the Majorana end-states operators with momentum $k$ and z-component of spin $\sigma$, take the form \cite{KitaevWire,Supplement}:  
\begin{align} \gamma_{{k}\sigma} & \approx \sum_{y} \phi_{k}(y) \[c_{{k}\sigma}(y)+i\sigma\text{sgn}(k)c_{-{k},-\sigma}^\dagger(y)\]
\end{align}
The states at $\pm k$ are not independent, but rather are related by: $\gamma_{k} = -\gamma_{-k}^\dagger \sigma_y$, where $\gamma_k = \begin{pmatrix} \gamma_{k\up} & \gamma_{k\down} \end{pmatrix}^T$.

Detailed expressions for the wave-function $\phi(y)$ are given in \cite{KitaevWire} (see also Appendix A).
For our present purposes, the only important feature of these wave-functions is their spatial-extent, and it is sufficient to use the approximate form $\phi_{k}(y) \approx\frac{e^{-y/\xi_k}}{\sqrt{\xi_k/2}}$.
The confinement length of the edge wave-functions depends on $k$, diverging as $\xi_k \approx \frac{\xi_0}{|k|}$ and $\xi_k\approx\frac{\xi_0^{-1}}{|\pm\Lambda-k|}$ near $k\approx 0,\pm\Lambda$ respectively, and falling to a minimum of $\xi_k\approx\frac{\xi_0}{\Lambda}$ near $k\approx\Lambda(1-\xi_0^{-2})$.  Here $\xi_0 = \frac{t}{\Delta_0}$ is the bulk coherence length.

\vspace{6pt}\noindent\textbf{\textit{Instability Towards FM - }}
We incorporate interactions by an on-site repulsive Hubbard term: $H_U = \frac{U}{2}\sum_i n_i(n_i-1)$ where $n_i$ is the number of electrons on site $i$.  First, we focus only on the sub-space of zero-energy edge states to analyze the possible competing instabilities.  From this approach, we identify the dominant tendency towards FM.  Next, we support this picture by numerically conducting a mean-field analysis that incorporates both surface and bulk states.  Due to the absence of quantum fluctuations about the Ferromagnetic ground-state we expect the mean-field description to be sufficient at low-temperatures, where thermal fluctuations are not important.

Spontaneous symmetry breaking order can endow the flat Majorana edge-bands with dispersion.  Focusing only on types of order that preserve translation symmetry, the generic edge dispersion takes the form:
\begin{align} 
\label{eq:MassTerms} H_m = \sum_{k}\gamma^\dagger_{k}\[m_0(k)+\v{m}(k)\cdot\bs{\sigma}\]\gamma_{k} \end{align}
where $m_\mu\in\mathbb{R}$, and $\bs{\sigma}$ are spin Pauli matrices.

The transformation properties of $M_{ab} = m_0\delta_{ab}+\v{m}\cdot\bs{\sigma}_{ab}$ allow us to identify the physical meaning of the various terms in $H_m$.  Hermiticity requires $M=M^\dagger$.  In contrast, time reversal acts on $M$ as: $\mathcal{T}(M) = -M^\dagger$, indicating that $H_m\neq 0$ necessarily breaks TRS. Spin rotations, $\gamma\rightarrow U\gamma = e^{-i\bs{\theta}\cdot\sigma}\gamma$ with $U\in \text{SU}(2)$ have the effect $M\rightarrow U^\dagger M U$.  Finally, under spatial inversion $x\rightarrow -x$, $M$, transforms acts like: $\mathcal{P}(M)= -\sigma^y M^T\sigma^y$.  These considerations show that $m_0$ and $\v{m}$ break inversion and spin-rotation symmetry respectively, allowing us to identify $m_0\neq 0$ with the $d+is$ wave pairing, and $\v{m}\neq 0 $ with edge FM (see Appendix C for an explicit derivation).

We write the Hubbard U term as: $H_U = -\frac{U}{2}\sum_{\v{r}}\[\(n_{\up,\v{r}}-n_{\down,\v{r}}\)^2-\(n_{\up,\v{r}}+n_{\down,\v{r}}\)\]$, and focus on configurations that are uniform along x. Decomposing the spin-density purely in terms of the edge states: $\sum_x\(n_{\up,\v{r}}-n_{\down,\v{r}}\) \approx \sum_{k,y}|\phi_{k}(y)|^2\gamma^\dagger_{k}\sigma^z\gamma_{k} +\dots$, where $\(\dots\)$ denotes the remaining contributions from bulk states gives:
\begin{align} H_U&\approx -\sum_{k,k'} V_{kk'}\(\gamma^\dagger\sigma^z\gamma\)_k\(\gamma^\dagger\sigma^z\gamma\)_{k'} +\dots
\end{align}
where $V_{kk'}\approx\frac{U}{2}\sum_{y}|\phi_k(y)|^2|\phi_{k'}(y)|^2
\approx\frac{U}{\(\xi_k+\xi_{k'}\)}$.
Here we see that it is energetically favorable to have $\<\gamma^\dagger\sigma^z\gamma\>\neq 0$, i.e. to magnetically polarize the edge.  A similar analysis starting by re-writing $H_U$ in terms of the s-wave pairing order parameter, $H_U\approx +\sum_{kk'}V_{kk'}(\gamma^\dagger\gamma)_k,(\gamma^\dagger\gamma)_{k'}$ shows, as expected, that it is energetically costly to form s-wave pairing at the edge.

Further insight into the ferromagnetic edge-instability can be gained by developing a Ginzburg-Landau free energy for the edge-magnetization.  Ignoring, for the moment, the bulk states, the effective action for the edge-Majorana bands (in imaginary frequency) is: $S_\gamma = \frac{1}{2}\sum_{\omega,k}\gamma^\dagger_{\omega,k}\(-i\omega\)\gamma_{\omega,k}+ H_U$.  Introducing the Hubbard-Stratonivich field $m_k\sim \sum_{k'}V_{kk'}\(\gamma^\dagger\sigma^z\gamma\)_{k'}$ to decompose the quartic term, and integrating out the Majorana edge-states, we find (see Appendix D): 
\begin{align} S_\text{eff} = \frac{1}{2}\sum_{kk'}m_k\(V^{-1}\)_{kk'}m_{k'}-\sum_{k}|m_k|\end{align}  The unusual, non-analytic $|m_k|$ term arises from the edge-bands' singular density of states in the absence of FM order.  This term favors $m\neq 0$, even for arbitrarily weak interactions.  The saddle point solution $\frac{\delta S_\text{eff}}{\delta m_k}=0$ is:
$m_k = \sum_{k'}V_{kk'}$.  The edge state dispersion is:
\begin{align}m_k\approx \frac{U}{\pi\xi_0}|k|\[\Lambda+|k|\log\(\frac{|k|}{\Lambda+|k|}\)\]
\end{align}

The above estimates give: $m_{k_\text{max}}\approx \Delta_0$. However, this is an overestimate for several reasons.  First, disorder will extrinsically broaden the surface-bands, smearing out the singular density of states and weakening the instability towards FM.  Second, edge roughness will further suppress the pairing near the edge, increasing the confinement lengths $\xi_k$, and weakening the residual interactions compared to the clean case. 
Lastly, we have so far ignored the mixing between surface and bulk states induced by the magnetic order.  This is justified only for small splitting ($m_k\ll\Delta_k$).

For larger splitting, the dependence of the confinement length on the splitting must be self-consistently taken into account: $\xi_k\sim \frac{v_\perp(k)}{\Delta_k-m_k}$, limiting the effects of interactions for large $m_k$.  The bulk states repel the surface-bands, keeping them inside the gap, even for arbitrarily large $U$. These effects can be accounted for by numerically simulating the full bulk plus edge problem using Eq.~\ref{eq:Hdwave} in the presence of an onsite repulsive $H_U = U\sum_i n_i(n_i-1)\approx -\frac{U}{2}\sum_{i}m_ic^\dagger_i\sigma^zc_i$ subject to the self-consistency constraint $m_i = \<c^\dagger_{ia}\sigma^z_{ab}c_{ib}\>$ (note that we allow the mean-field parameters to have arbitrary spatial variation in the direction perpendicular to the edge).  Representative results using reasonable values of $t,U,\Delta_0$ are shown in Fig.~\ref{fig:EdgeSpectrum}.

\vspace{6pt}\noindent\textbf{\textit{Tunneling Signatures - }}
The edge bands and ferromagnetic splitting can be revealed by tunneling\cite{ZBP1,ZBP2,ZBP3,Matsumoto,Fogelstrom}.  At zero temperature, each edge-mode contributes a Lorentzian peak centered at bias $eV = \pm m_k$, with height $\frac{2e^2}{h}$, and width $\gamma_k\approx \frac{\gamma_0}{\xi_k}\approx \gamma_0|k|$.  Here, $\gamma_0 \equiv \pi\nu(0)|\Gamma|^2$ is the typical level-broadening due to coupling to the metallic lead, where $\nu(0)$ is the density of states of the metallic lead, and $\Gamma$ is the lead-superconductor tunneling amplitude.  We have calculated the detailed tunneling conductance based on a mean-field treatment of $H_U$ and Eq.~\ref{eq:Hdwave}.  The results, are in agreement with the analytic considerations presented above.

\begin{figure}[ttt]
\begin{center}
\includegraphics[width = 3.5in]{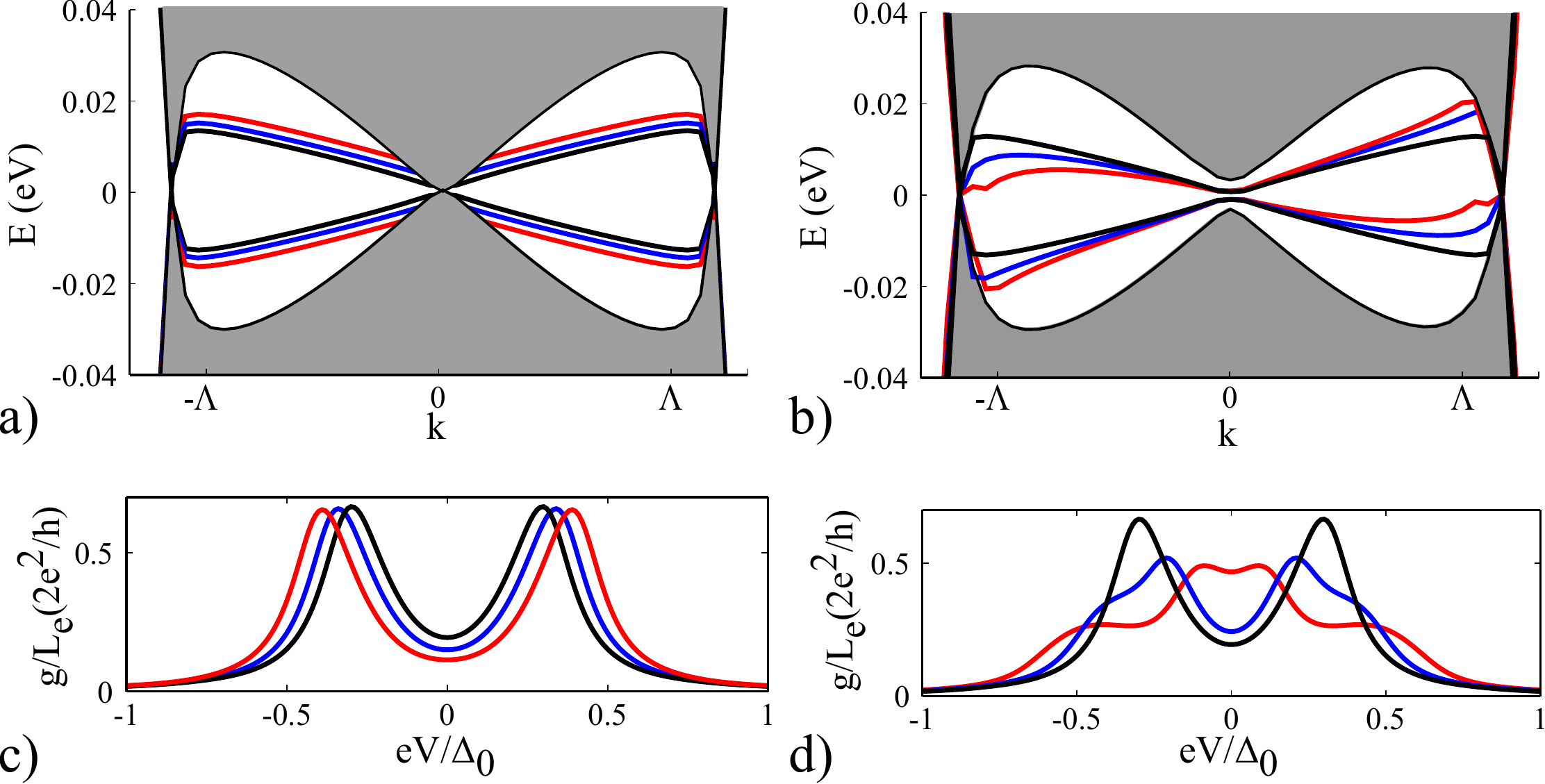}
\end{center}
\vspace{-.2in}
\caption{A magnetic field further splits the edge bands, and corresponding tunneling peak. Panels a) and b) show the edge-spectrum (solid lines) and projected-bulk-spectrum (gray shaded regions) for parallel and perpendicular magnetic field.  (c,d) show the corresponding tunneling conductance from the edge states, obtained from a numerical mean-field analysis.  In black, blue, and red curves correspond to $H_\parallel=$0, 7T, and 14T for (a,c), and to $H_\perp=$0, 100mT, and 200mT for (b,d). }
\vspace{-.2in}
\label{fig:FieldSplitting}
\end{figure} 

The tunneling conductance from the edge bands, normalized to the length of the edge is shown in Fig.~\ref{fig:EdgeSpectrum}c.  For weak tunneling, $\gamma_0\ll 1$, the edge-state contribution to the tunneling conductance at bias voltage $V$ is given by: 
\begin{align} g(eV) = \frac{2e^2}{h}\frac{\gamma_0L_e}{ 2\xi_0}\frac{k(eV)}{v_e(eV)}\end{align}
where $L_e$ is the length of the edge, $k(\e)$ is the momentum of the edge state with splitting $\e$ (defined by $m_{k(\e)}=\e$), and $v_e(\e) = |\frac{\partial m_k}{\partial k}|_{m_k=\e}$ is the corresponding edge velocity.  The conductance exhibits a peak near the maximal energy splitting, $m_{k_\text{max}}$, where $v_e$ flattens out providing large edge-mode density of states that have comparatively large weight near at edge.  The width of this peak is proportional to $\gamma_0$.  

\vspace{6pt}\noindent\textbf{\textit{Parallel Magnetic Field - }}  An applied magnetic field $\v{H}$ induces further TRS breaking perturbations, and further splits the edge states (see Appendix F for more details).  We first consider the simpler case of an in-plane magnetic field, which affects the edge states through the Zeeman coupling: $\mathcal{H}_z\approx \frac{g\mu_B}{2}\v{H}\cdot\sum_{k}\gamma^\dagger_{k}\bs{\sigma}\gamma_{k}$.
In the FM scenario described above, this Zeeman energy simply adds to the spontaneous zero-field splitting, further splitting the edge-bands and tunneling peak (see Fig.~\ref{fig:FieldSplitting}a,c).  In fact, such a splitting of the tunneling peak with an in-plane field, with slope equal to the Zeeman splitting has been observed\cite{Krupke}.

In contrast, no such splitting is expected for the previously suggested scenario in which the zero-field splitting is due to $d+is$ pairing near the edge\cite{Matsumoto,Fogelstrom}.  Spontaneous $d+is$ edge pairing corresponds to an $m_0$ type mass (see Eq.~\ref{eq:MassTerms}) at zero-field.  Applying an in-plane field induces an $\v{m}$ type term, which further splits the tunneling peak into four.  However, the further splitting occurs symmetrically about the zero-field peak position, and there will be no shift in the average peak position.  Consequently, the in-plane field data of \cite{Krupke} provides evidence for spontaneous FM rather than a sub-dominant TR breaking pairing.

\vspace{6pt}\noindent\textbf{\textit{Perpendicular Magnetic Field - }} In addition to Zeeman effects, a magnetic field, $H_\perp$, perpendicular to the ab-plane will produce orbital screening currents.  These introduce the edge perturbation: $\mathcal{H}_A = \int \v{A}\cdot\v{j} \approx \sum_k m_0(k)\gamma^\dagger_{k}\gamma_k$.  Here $\v{A} = \lambda_LHe^{-y/\lambda_L}\hat{x}$ is the vector potential corresponding to uniform perpendicular field $H$ (in the unitary gauge), and $m_0(k) = \int dy A(y)|\phi_k(y)|^2\approx \frac{H}{H_c}\Delta(k)$.  $\lambda_L$ is the London penetration depth, $H_c = \frac{\Phi_0}{\pi\xi_0\lambda_L}$ is the thermodynamic critical field, and $\Phi_0 = \frac{hc}{2e}$ is the superconducting flux quantum.  For YBCO, $H_c\approx 1T$.  Note that, in the presence of vortices, $\v{A}$ should be replaced by $\v{A}-\frac{c}{2e}\nabla\theta$ in the above expression, where $\theta$ is the superconducting phase.

\begin{figure}[ttt]
\begin{center}
\includegraphics[width = 3.2in]{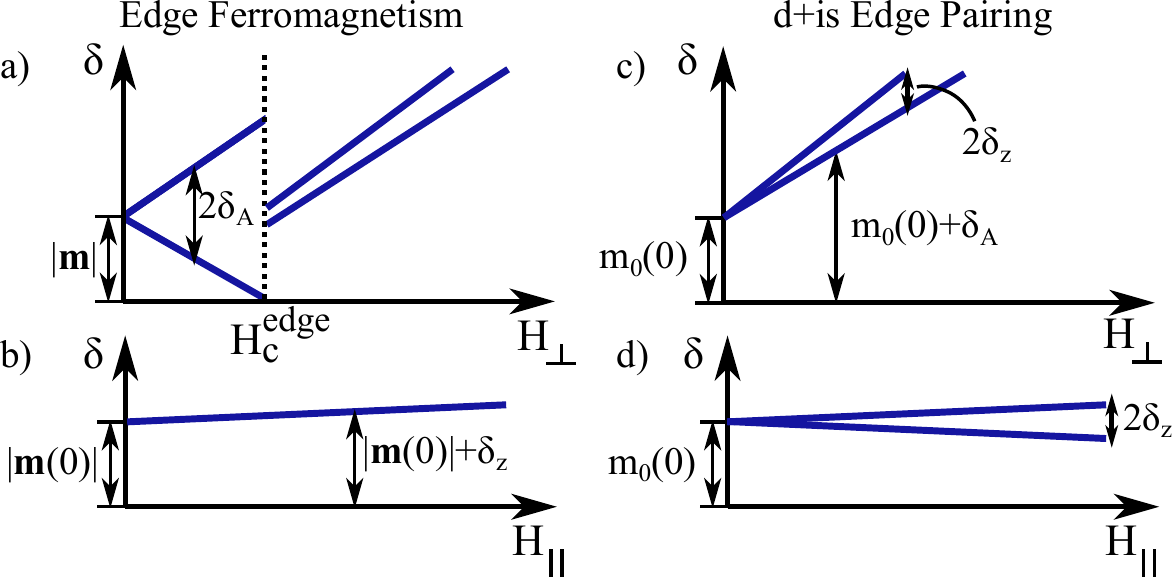}
\end{center}
\vspace{-.2in}
\caption{Schematic dependence of tunneling peak position, $\delta$, on magnetic field, for edge-FM (panels a,b) and $d+is$ (panels c,d) scenarios.  Only the positive bias peak is shown here (negative bias is symmetric).  The top (a,c) and bottom (b,d) rows show the effects of perpendicular and parallel fields respectively.  The dashed line denotes a first-order phase transition where the spontaneous edge-FM is destroyed by the applied field.  $\delta_Z$ and $\delta_A$ are respectively the splittings induced by the Zeeman energy and orbital screening currents from the external field.
}
\vspace{-.2in}
\label{fig:SchematicFieldSplitting}
\end{figure}

For $H<H_c$, the screening currents from $H_\perp$, induce an $m_0$-type term that, in the FM scenario, shifts the $k>0$ and $k<0$ edge-state energies in opposite directions (see Fig.~\ref{fig:FieldSplitting}c), leading to a four-fold split tunneling peak (see Fig.~\ref{fig:FieldSplitting}d).  As $H$ is increased beyond $H_c^\text{edge}\approx 250mT$, the negative energy spin-down states are pulled above the chemical potential, and the spontaneous edge FM is killed in a first-order field-induced phase transition.  Past this point, the tunneling conductance shows only a two-fold split tunneling peak, centered at $eV\approx \pm \frac{H}{H_c}\Delta_0$.  Simulated tunneling conductance from a mean-field treatment of Eq.~\ref{eq:Hdwave} with $H_U$ and $\mathcal{H}_{A,Z}$ agree with the schematic picture presented in Fig.~\ref{fig:FieldSplitting}.  In addition, the perpendicular field simulations show a very broad peak for $H>H_c^\text{edge}$ (see Appendix F).

In contrast, for the previously suggested $d+is$ scenario, the orbital currents simply add to the zero-field splitting.  Therefore, in principle, the effects of a perpendicular field on tunneling conductance can provide further evidence to distinguish the FM and $d+is$ scenarios.
There is insufficient data from \cite{CovingtonSplitting,Krupke} at low fields, to discern whether there is first order transition in perpendicular field at $H_\perp=H_c^\text{edge}$.  It is also possible that our estimate of $H_c^\text{edge}$ is inaccurate due to vortex pinning effects.  There are other complications in the data which are not accounted for in our simple model.  The peak shift saturates at large $H_\perp$, however the previously proposed explanation based on non-linear saturation of the Meissner currents\cite{Fogelstrom,Yip} is also applicable to our scenario.  There are also complicated hysteresis effects in perpendicular field which are not easy to explain\cite{Endnote:Hysteresis}.  For these reasons, we suggest that the parallel field dependence provide simpler, and more easily interpreted evidence.

\vspace{6pt}\noindent\textbf{\textit{Discussion - }}  We have shown that the topological flat-bands provide a useful perspective for discussing TRS breaking at the edge of high-Tc cuprate superconductors.  This viewpoint  naturally suggests an instability towards ferromagnetic order from repulsive interactions.  The magnetic-field dependence of tunneling peak splitting provides a simple, though indirect, experimental test to distinguish the proposed FM order from previously proposed $d+is$ pairing\cite{Matsumoto,Fogelstrom}.  Alternatively spin-polarized planar-tunneling or STM measurements would enable one to directly detect of the predicted FM order.

In closing, we remark that flat Majorana bands are also expected to appear in certain classes of nodal spin-triplet superconductors\cite{Wong}.  In contrast to the d-wave case discussed here, these edge bands would not be spin-degenerate, and more complicated types of (incommensurate density wave) order are required.  Moreover, unlike the ferromagnetic case discussed here, quantum fluctuations would generically destroy the density-wave order leaving only power-law correlations.

\textit{Acknowledgements - } We thank Liang Fu for helpful discussions, and acknowledge funding from DOE Grant No. DEFG0203ER46076.

\newpage
\appendix

\section{Appendix A. Edge State Wave-functions }
As shown in \cite{KitaevWire}, the edge-state zero-mode wave-function at fixed k along the edge is:
\begin{align} \phi_k(y) &= \sum_{y}\frac{1}{N_k}\(\lambda_+(k)^y+\lambda_-(k)^y\) 
\nonumber \\
\lambda_\pm &= \cfrac{-\mu\pm\sqrt{\mu^2+4\Delta(k)^2-4t(k)^2}}{2\(|t(k)|+|\Delta(k)|\)}
\end{align}
where $N_k = \sum_y|\lambda_+(k)^y+\lambda_-(k)^y|^2$ ensures proper normalization. 

\vspace{.1in}\noindent\textbf{\small A.1 - Divergence of $\xi_k$ Near Bulk Nodes }\\ \indent 
Near $k\approx 0$, $\Delta_k\rightarrow 0$, $\lambda_\pm\approx\frac{-\mu\pm i\sqrt{4t_k^2-\mu^2}}{2t_k}$, i.e. $|\lambda_\pm| = 1$.  Since the wave function decays as $e^{y\ln\lambda}$, this indicates a diverging confinement length.  Expanding near $\Delta\approx 0$ gives:
\begin{align} |\lambda_\pm|^2\approx \frac{4t_k^2-4\Delta_k^2}{2(t_k+\Delta_k)}\approx 1-\frac{2\Delta_k}{t_k} \end{align}
or $\xi_k^{-1}=-\ln|\lambda_\pm|\approx \frac{|\Delta_k|}{t_k}\approx \frac{\Delta_0}{t_0}|k|$ i.e.:
\begin{align} \xi_{k\approx 0} \approx \frac{\xi_0}{|k|} \end{align}
where $\xi_0 = t_0/\Delta_0$ is the bulk coherence length.

Expanding near $|k|\approx \Lambda$, on the other hand, we denote $t_k = \mu/2+\delta t_k$ with $\delta t>0$, then $\lambda_+\rightarrow \frac{-\mu+2\Delta_k}{\mu+2\Delta_k}$ is regular as $\delta t\rightarrow 0$.  The diverging binding-length therefore comes from $|\lambda_-|\rightarrow 1$:
\begin{align} -\lambda_- \approx \cfrac{\mu+2\sqrt{\Delta^2-\mu\delta t}}{\mu+2\Delta+2\delta t} \approx 1-\frac{\delta t}{\Delta}
\end{align}
$\lambda_+^y$ goes to zero rapidly for small $y$, the wave-function can be approximate as $|\lambda_-^y| = e^{y\ln\lambda_-}\approx e^{y\ln(1-\delta t/\Delta)}\approx e^{-y\delta t/\Delta}$.  Near $k\approx \pm \Lambda$, $\delta t_k \approx 2t\sin(k)\delta k$, and $\Delta_k \approx 2\Delta_0\sin(k)$.  The confinement length in this limit is then:
\begin{align} \xi^{-1}_{|k|\approx\Lambda}\approx \frac{t}{\Delta_0}\(\Lambda-|k|\)\end{align}

\vspace{.1in}\noindent\textbf{\small A.2 - Interpolating Expression}\\ \indent 
More generally, $\xi_k\approx \frac{v_{F,\perp}}{E_\text{min}(k)}$ where $v_{F,\perp}(k) = \frac{\partial E(k,k_\perp)}{\partial_k}|_{k_\perp=k_{F\perp}}$, and $E_\text{min}(k) = \min_{k_\perp} E(k,k_\perp)$, which is the natural coherence length scale for the effective 1D chain at momentum $k$.  

Near $k\approx \pm\Lambda$, $v_{F,\perp} \approx v_\Delta = 2\Delta_0$, and $E_\text{min}\approx v_{F,\parallel} = v_F = \sqrt{4t^2-\mu^2}\approx  2t$ (for small doping).  Conversely, near $k\approx 0$, then $v_{F,\perp}\approx v_F$, and $v_{F,\parallel}\approx v_\Delta$.

Interpolating between these two limits, we expect a minimum confinement length of $\(\xi_k\)_\text{min}\approx \frac{1}{\xi_0\Lambda}$ near $k\approx \pm\(1-\frac{1}{\xi_0^2}\)\Lambda$.  Since typically, $\xi_0\gg 1$, (see Fig.~\ref{fig:EdgeSpectrum}), there are very few states in the region $\(1-\frac{1}{\xi_0^2}\)\Lambda<|k|<\Lambda$.  These few states make only a small contribution to the energetics of the edge, and for most purposes it is sufficient to approximate $\xi_k$ by the monotonic function:
\begin{align}\xi_k\approx \frac{\xi_0}{|k|} \end{align}

\section{Appendix B.  Action of Symmetries on Edge-Splitting Terms}
Using the convention $k>0$, and $\Delta_0$ real throughout, the zero-mode wave-functions can be schematically written as:
\begin{align}
\gamma_{k\up} &\sim c_{k\up}+ic_{-k\down}^\dagger 
\nonumber\\
\gamma_{k\down} &\sim c_{k\down}-ic_{-k\up}^\dagger 
\nonumber\\
\gamma_{-k\up} &\sim c_{-k\up}-ic_{k\down}^\dagger 
\nonumber\\
\gamma_{-k\down} &\sim c_{-k\down}+ic_{k\up}^\dagger
\end{align}

\vspace{0.2in}\noindent\textbf{\small B.1 - Spin Rotation}\\
Consider a generic (spatially uniform along x) mass term:
\begin{align} \sum_{k>0}\gamma_{ka}\tilde{M}_{ab}(k)\gamma_{-kb} \end{align}
where, $a,b\in\{\up,\down\}$ label spin.
Note, we could also extend our summation interval to include $k<0$, but in this case one can easily see (by changing variables $k\rightarrow -k$ and then anti-commuting the $\gamma$'s) that only the components satisfying $\tilde M(-k) = -\tilde M^T(k)$ survive the summation; consequently the $\pm k$ contributions are not independent.

Since, $\gamma_{\pm ka}$ transform like conventional spinors under spin-rotation, we see that under $\gamma\rightarrow e^{i\bs{\theta}\cdot\bs{\sigma}}\gamma \equiv U_{\bs{\theta}}\gamma$, we have: $M\rightarrow U_{\bs{\theta}}^T M U_{\bs{\theta}} = \sigma^y U^{-1}\sigma^yMU$.  This transformation suggests that we parameterize $\tilde{M}=\sigma^yM$, such that $M$ transforms as a $(1/2,1/2)$ tensor under spin-rotations.  

\vspace{0.2in}\noindent\textbf{\small B.2 - Hermitian Conjugation}\\
Under Hermitian conjugation the edge states transform as:
\begin{align} \gamma_{ka}^\dagger = \sigma^y_{ab}\gamma_{-kb} \end{align}
Correspondingly, the edge-splitting terms transform like:
\begin{align} \(\gamma_{ka}\tilde{M}_{ab}(k)\gamma_{-kb}\)^\dagger & = \tilde{M}^*_{ab}(k)\gamma_{-kb}^\dagger\gamma_{ka}^\dagger 
\nonumber\\
&=\tilde{M}^*_{ab}(k) \(-\sigma^y_{bb'}\gamma_{kb'}\)\(\sigma^y_{aa'}\gamma_{-ka'}\) 
\nonumber\\
&=\gamma_{ka}\[\sigma^y \tilde{M}^\dagger \sigma^y\]_{ab}\gamma_{-kb}
\end{align}
This constrains $M = M^\dagger$, which can be accomplished by parameterizing $M = m_0+\v{m}\cdot\bs{\sigma}$.

\vspace{0.2in}\noindent\textbf{\small B.3 - Time-Reversal}\\
Time-reversal (TR) acts on the electron operators like $c_{ka}\rightarrow i\sigma_yKc_{kb}$ where $K$ represents complex conjugation.  In terms of the edge modes, one then has $\gamma_{ka}\rightarrow i\sigma^y_{ab}\gamma_{-kb}$, and:
\begin{align} \(\gamma_{ka}\tilde{M}_{ab}(k)\gamma_{-kb}\) &\rightarrow i\sigma^y_{aa'}\gamma_{-ka'}\tilde{M}^*_{ab}i\sigma^y_{bb'}\gamma_{kb'} 
\nonumber\\
&= \gamma_{kb'}\(\sigma^y\)^T_{b'b}M^\dagger_{ba}\sigma^y_{aa'}\gamma_{-ka'}
\end{align}
or equivalently:
\begin{align}M\rightarrow -M^\dagger \end{align}
We see that the constraint of Hermiticity necessarily requires that any mass term break TR symmetry.  

\vspace{0.2in}\noindent\textbf{\small B.4 - Spatial Inversion}\\
Inverting the x-direction acts projectively in the presence of d-wave pairing.  In order to compensate for the sign change of the order parameter under $x\rightarrow -x$, we must accompany this inversion by a global gauge transformation $c\rightarrow \pm i c$.  Since the particular sign is irrelevant, let us choose $c\rightarrow ic$ for definiteness, in which case we have $\gamma_{\pm k \sigma} \rightarrow i\gamma_{\mp k\sigma}$.  Note that, in contrast, $d+is$ pairing transforms to $d-is$ pairing under this transformation.  It is not possible to relate $d\pm is$ pairing by a global gauge transformation as the relative sign of the $d$- and $s$-wave components is physically observable, corresponding to the direction of edge currents from the TRS breaking pairing.

The mass term then transforms as:
\begin{align} M\rightarrow -\sigma^yM^T\sigma^y \end{align}
Using the parameterization $M = m_0+\v{m}\cdot\bs{\sigma}$, $m_0\rightarrow -m_0\mathbb{I}$ and $\v{m}\rightarrow\v{m}$.

\section{Appendix C. Explicit Representation of Electron Bilinears by Edge States}
Decomposing the electron plane-wave operators $c_{k\sigma}$ into the single particle eigenstates we have: \begin{align}
c_{k\up}^\dagger(y) &= -i\phi_k(y)\gamma_{-k\down}+\dots \nonumber\\
c_{k\up}(y)  &=\phi_k(y)\gamma_{k\up}+\dots \nonumber\\
c_{k\down}^\dagger(y) &= i\phi_k(y)\gamma_{-k\up}+\dots \nonumber\\
c_{k\down}(y) &= \phi_k(y)\gamma_{k\down}+\dots
\end{align}
where we have explicitly written only the contribution from the zero-mode edge states, and $(\dots)$ indicate contributions from other extended states, which we ignore for this section.  Similarly, we may decompose the spin-density and s-wave pairing in terms of single-particle eigenstates.  Retaining only the contributions from the zero-energy edge states we have:
\begin{align} S^z(q=0) &=\sum_{k}\[c^\dagger_{k\up}c_{k\up}-c^\dagger_{k\down}c_{k\down}\]
\nonumber\\
&\approx \sum_{k}|\phi_k|^2\[-i\gamma_{-k\down}\gamma_{k\up}- \(i\gamma_{-k\up}\gamma_{k\down}\) \]+\dots
\nonumber\\
&=\sum_{k}|\phi_k|^2\[i\gamma_{k\up}\gamma_{-k\down}+i\gamma_{k\down}\gamma_{-k\up} \]+\dots
\nonumber\\
&= \sum_k|\phi_k|^2 \gamma^\dagger_k\sigma^z\gamma_{k} +\dots
\end{align}

Similar, the s-wave pairing density is:
\begin{align} \mathcal{F}_s &= \sum_k \(c_{k\up}^\dagger c_{-k\down}^\dagger +c_{-k\down}c_{k\up}\) 
\nonumber\\
&\approx \sum_k \phi_k^2 \[\(-i\gamma_{-k\down}\)\(i\gamma_{k\up}\)+\(\gamma_{-k\down}\)\(\gamma_{k\up}\)\]
+\dots
\nonumber\\
&= -\sum_k\phi_k^2 \gamma^\dagger_k\mathbb{I}\gamma_{k}+\dots
\end{align}

Note that, for repulsive on-site interactions $Un_i(n_i-1)$ can be decomposed as $-US_z^iS_z^i$ or $U|\mathcal{F}_s|^2$, indicating that the $\v{m}$ terms is energetically favored by repulsive interactions ($U>0$), whereas the $m_0$ mass is not.

\section{Appendix D. Derivation of Effective Landau-Ginzburg Action}
Consider just the edge modes, then the effective action is:
\begin{align} \label{eq:Sgamma} S = \sum_k \gamma^\dagger\partial_\tau\gamma - \frac{1}{2}\sum_{kk'}V_{kk'}\(\gamma^\dagger\sigma^z\gamma\)_k\(\gamma^\dagger\sigma^z\gamma\)_{k'}
\end{align}
Applying Hubbard-Stratonovich to decouple $U$ term:
\begin{align} S = \frac{1}{2}\sum_{kk'}m_kV^{-1}_{kk'}m_k'+ \sum_{\omega,k}\gamma^\dagger\(-i\omega+m_k\sigma^z\)\gamma \end{align}

Integrating out $\gamma$'s gives a term:
\begin{align} S_{\gamma}=\text{tr}\ln\[i\omega-m_k\sigma^z\] = -\sum_{\omega,k}\ln\[\omega^2+m_k^2\]+\text{const} \end{align}

The $\omega$ integration can be performed by subtracting an $m$ independent constant to regulate the $\omega\rightarrow \infty$ behavior:
$\sum_{\omega}\ln\(\omega^2+m_k^2\) \rightarrow\sum_{\omega}\ln\frac{\omega^2+m_k^2}{\omega^2+\delta^2} $
Here, $\delta$ arbitrary (for concreteness, we take $\delta<m$), and we are only interested in the $m$ dependence of the final answer (since m-independent terms will not affect the saddle-point solution).  The term $\ln\(\omega^2+m^2\)$ has a branch cut, which can be taken between $\omega = ix$, $x\in [-|m|,|m|]$.  Similarly, the $\ln\(\omega^2+\delta^2\)$ branch cut can be taken between $[-i\delta,i\delta]$.  The branch cut of the two logs cancel between $\omega \in [-i\delta,i\delta]$ when taken together as $\ln\frac{\omega^2+m_k^2}{\omega^2+\delta^2}$, leaving the integrand well defined on the real axis.  Moreover, since as $|\omega|\rightarrow \infty$, the integrand falls off as $\sim \frac{1}{\omega^2}$, we may extend the integration range to a contour encircling the full upper half-plane.  Since, in the upper half-plane (UHP), the integrand is analytic everywhere outside $\omega \in [i\delta,im]$ on the imaginary axis, the contour can be shrunk to one encircling (but infinitesimally outside) the branch-cut in the UHP.  On each side of the branch cut, the integrand differs by $-2\pi i$ (for $\omega = xe^{i(\pi/2\pm \epsilon)}$, with $x\in[\delta,m]$, then as $\epsilon\rightarrow 0$ $\ln\(\omega^2+m^2\)$ is independent of $\epsilon$ but $-\ln\(\omega^2+\delta^2\)$ differs by $-2\pi i$ between $+\epsilon$ and $-\epsilon$). Consequently the integral reduces to $-2\pi i\int_{i\delta}^{i|m|}\dslash \omega = |m|-\delta$  Consequently we find:
\begin{align} S_\gamma = -\sum_k |m_k|+\text{const} \end{align}

A less direct, but much simpler way to evaluate the sum is to note that that $\frac{\delta S_\gamma}{\delta m_k} = \sum_{\omega} \frac{2m_k}{\omega^2+m_k^2} = \text{sgn}(m_k)$, which indicates that $S_\gamma = -\sum_k |m_k|+\text{const}$.

For repulsive interactions (i.e. $V_{kk'}>0$), we can choose $m>0$, and drop the absolute value signs.  Therefore, we see that in the free energy for $m$, there is a linear term due to the singular density of states of the flat bands in the $m\rightarrow 0 $ limit.  The resulting saddle point solution and energy are simply:
\begin{align} m_k|_\text{sp} = \sum_{k'}V_{kk'} \hspace{.25in} 
S|_\text{sp} = -\frac{1}{2}\sum_{kk'}V_{kk'}\end{align}

This consideration is useful, because, generically the bulk will add a term that suppresses $m_k$, which we can write schematically as $\sum r|m_k|^2$.  However, due to the linear term from the edge states, the saddle point will always have non-zero $m$, and energy that tends as $\sim -\frac{\sum_{kk'}V^{-1}_{kk'}}{r^2}$ even when magnetization is arbitrarily strongly disfavored in the bulk ($r\rightarrow \infty$).

\vspace{.2in}\noindent\textbf{\small D.1 - Mean-Field Energy and Dispersion}\\
The end-state wave-functions may be reasonably approximated by:
$\phi_k(y) \approx \frac{2}{\sqrt{\xi_k}}e^{-y/\xi_k}$.
This ignores the oscillatory structure, which is reasonable when calculating matrix elements like $\int dy |\phi_k(y)|^2|\phi_{k'}(y)|^2$ that will appear below, since, for $k$ very different from $k'$, the oscillations in $\phi_k$ and $\phi_{k'}$ happen with very different periods, in which case $|\phi_k|^2$ sees roughly the average of $|\phi_{k'}|^2$.

At zero-temperature, approximating $\xi_k\approx \xi_0/|k|$, the 
The dispersion $m_k$ is approximately given by:
\begin{align}  m_k &= \sum_{k'}V_{kk'}\approx\frac{2U}{\xi_0}\int_0^\Lambda\dslash k'\frac{kk'}{k+k'} \nonumber\\&=\frac{U}{\pi\xi_0}k\(\Lambda+k\log\(\frac{k}{k+\Lambda}\)\) \end{align}
This is strictly only valid near $k\approx 0$, and that as $k$ becomes very close to $\pm\Lambda$, the spectrum should dip back to zero.  To capture this behavior, we should replace: $k\rightarrow \xi_0^2\(\Lambda-|k|\)$ in the above expression.  However, for energetic purposes, most of the states have dispersion following that described above. 

Similarly, the mean-field energy gain is:
\begin{align} E_\text{MF} &= \sum_{kk'}V_{kk'}=
\frac{U}{2\pi^2\xi_0}\int_0^\Lambda dk k\(\Lambda+k\log\(\frac{k}{k+\Lambda}\)\)
\nonumber\\
&= \(\frac{\log\frac{e}{2}}{3\pi^2}\)\frac{U\Lambda^2}{\xi_0} \approx 0.01\times \frac{U\Lambda^2}{\xi_0}\end{align}

\section{Appendix E. Details of Mean-Field Simulations}
Mean-field simulations of the tight-binding model Eq.~\ref{eq:Hdwave} of the main text, with an on-site Hubbard U term were conducted numerically.  We decomposed the Hubbard $U$ interaction in terms of spin, and considered magnetization profiles that uniform along the edge (x-direction), but spatially inhomogeneous along $y$.  In this case, translational symmetry was preserved along $x$, and the states at different $k$ were decoupled, enabling the problem to be reduced to a collection of separate 1D models with the same magnetization profile.  The simulations displayed in the main text discretized $k$ into 51 values (the results did not change markedly for larger numbers of $k$), and the system length in the y-direction was $L_y=900$ sites (further increasing the length led only to minimal changes).  Self-consistency was achieved by starting with an initial ansatz for the magnetization, and iteratively computing the induced mean-field magnetization until the fractional (root-mean-squared) change in magnetization between iterations was $<1\%$.

Tunneling conductance was computed as $\frac{2e^2}{h}\text{tr} |r_{eh}|^2$, where $r_{eh}$ is the reflection matrix for processes in which an incoming electron from the lead is Andreev reflected as a hole.  We assumed a uniform coupling to between the lead and the wire, that is independent of the momentum along the edge.  This is appropriate for a thin, high tunneling barrier.  Since the effective barrier thickness depends on the incident angle of the incoming electron, for a thicker barrier, the lead-superconductor coupling will depend on the incident angle of the incoming electron.  This will obscure the underlying behavior of the tunneling peak by providing a strongly momentum dependent visibility to the edge-states.

In the next appendix, we provide further details about the impact of perpendicular magnetic fields, and incorporate orbital magnetic field effects into the numerical mean-field simulations.

\section{Appendix F. Orbital Magnetic Field Effects}

This appendix further details on the orbital effects of a perpendicular field.  For $H<H_{c1} = \frac{\Phi_0}{\pi\Lambda_L^2}\approx 30$mT, out of plane fields are screened by the superconductor as:
\begin{align} H(y)\hat{z} = H\hat{z}e^{-y/\lambda_L} \end{align}
where $\lambda_L$ is the London penetration depth ($\lambda_L\gg \xi_0$ for cuprates). For $H>H_{c1}$, vortices enter the system and the current profile at the edge depends in detail on the precise distribution of vortices, which will depend on pinning effects from the sample edge and from impurities.  A realistic description of perpendicular field effects in actual cuprate edge-tunneling experiments would need to confront this  complication.  However, it is useful to gain some intuition from the simpler case of uniform edge currents, and we will imagine throughout that the field is below $H_{c1}$ (even though we consider fields that would exceed the $H_{c1}$ for realistic cuprate materials).

Working in the unitary gauge (where the superconducting phase is fixed uniformly to zero), $H$ can be parameterized by the corresponding vector potential:
\begin{align} \v{A} = H\lambda_Le^{-y/\lambda_L}\hat{x} \end{align}
where $\lambda_L$ is the bulk penetration depth.  This induces the perturbation (that we re-write in terms of the edge states):
\begin{align} \int \v{A}\cdot\v{j} &= e\sum_{k}\int dy A_x(y)v_x(k)c^\dagger_{ky}c_{ky} 
\nonumber\\
&\approx e\sum_{k}\(\int dy B\lambda_Le^{-y/\lambda_L}|\phi_k(y)|^2\)v_x(k)\gamma^\dagger_{k}\gamma_{k} 
\nonumber\\
&\approx eB\lambda_Lv_F\sum_{k}\frac{|k|}{\Lambda}\gamma^\dagger_{k}\gamma_{k}
\approx \frac{\pi B\lambda_L\xi_0}{\Phi_0}\sum_{k} \Delta(k)\gamma^\dagger_{k}\gamma_{k}
\nonumber\\
&\approx \frac{B}{ H_c}\sum_{k}\Delta(k)\gamma^\dagger_{k}\gamma_{k}\end{align}
where in the first line, we have used $\xi_0\ll \lambda_L$, which means that for most $k$, $\xi_k\ll \lambda_L$ (except those very close to the bulk nodes, which contribute only weakly to tunneling conductance).  In the last line, $\Phi_0 = \frac{hc}{2e} = \frac{\pi}{e}$ is the superconducting flux quantum, and we have introduced the critical field $H_c = \frac{\Phi_0}{\pi\xi_0\lambda_L}$ ($\approx 1T$ for YBCO).

\vspace{0.2in}\noindent\textbf{\small F.1 - Orbital Currents in the Vortex State}\\
In this section, we consider the effect of a finite density of vortices on the edge-current profile.  Generically, we find that the typical current density near the edge is unaffected by the presence of vortices.  

In the extreme type-II limit, valid for the quasi-2D high-Tc materials, the vortex cores do not strongly effect the current distribution at the edge.  Neglecting the contribution from vortex cores, the Landau-Ginzburg free energy density of the superconductor in a perpendicular external field, $H$ is:
\begin{align} \mathcal{F} = \frac{\rho_s}{4m}\(\nabla\theta-\frac{2e}{c}\v{A}\)^2 +\frac{1}{8\pi}\(\nabla\times\v{A}-H\hat{z}\)^2 \end{align}
where $\rho_s$ is the superfluid density, $\theta$ is the phase of the superconducting order parameter, and the $H\hat{z}$ term comes from external current sources that attempt to impose a constant magnetic field: $\v{B}=\v{H}$.

It is convenient to absorb the phase-gradient, $\nabla\theta$ into the vector potential: $\v{A}\rightarrow \v{\tilde{A}}=\v{A}+\frac{c}{2e}\nabla\theta$.  It is then $\v{\tilde{A}}$ which acts as the orbital perturbation for the edge states rather than $\v{A}$.  So long as we impose $\v{\tilde{A}}\cdot\hat{n}=0$, where $\hat{n}$ is the normal vector for the sample boundary, then we can interpret $\v{\tilde{A}}$ as being proportional to the supercurrent $\v{j}_s = -\frac{4\pi}{\lambda_L}\v{\tilde{A}}$.  It is crucial to recognize that it is $\v{\tilde{A}}$ and not $\v{A}$ which couples to the edge states.  We shall see that, while $\v{A}$ is strongly dependent on the vortex distriubtion, near the sample edge, $\v{\tilde{A}}$ is not.

With this transformation the free-energy density reads:
\begin{align} 4\pi\mathcal{F} &= \frac{1}{\lambda_L^2}|\v{\tilde{A}}|^2+ \frac{1}{2}\(\nabla\times\v{\tilde{A}}+\Phi_0\rho_v(\v{r})-H\)^2
\end{align}
where $\lambda_L$ is the London penetration depth, $\Phi_0 = \frac{hc}{2e}$ is the superconducting flux quantum and $\rho_v(\v{r}) = \frac{\hat{z}\cdot\nabla\times(\nabla\theta)}{2\pi}$ is the density of vortices.  Choosing $\v{\tilde{A}} = \tilde{A}_x(y)\hat{x}$, and minimizing with respect to $\v{\tilde{A}}$, gives the equation of motion:
\begin{align} \(-\partial_y^2+\frac{1}{\lambda_L^2}\)\tilde{A}_x(y)=-\Phi_0\partial_y\rho_v(y)
\end{align}

We approximate the vortex distribution by a uniform continuous density (valid in the limit where the inter-vortex spacing is much shorter than the $\lambda_L$, but much larger than the core size $\xi_0$). Specifically, we consider $\rho_v(\v{r}) = \frac{H}{\Phi_0}\theta(y-L)$ where $\theta$ is the step-function, and $L$ is some unknown distance, which can, for example, account for barrier effects repelling the vortices from the sample edge.

Solving the above equations, subject to the boundary condition that $-\partial_yA_x(y=0) = H$ gives:
\begin{align} \v{\tilde{A}} = H\lambda_L\hat{x}\[e^{-|y-L|/\lambda_L}+(1-e^{-L/\lambda_L})e^{-y/\lambda_L}\]
\end{align} 

Since the edge states reside predominately within distance $\xi_0\ll \lambda_L$ of the edge, they are effected by $\v{\tilde{A}}_x(y\rightarrow 0) = H\lambda_L$.  This result is independent of $L$, indicating that the typical size of the orbital perturbation is insensitive to the presence and distribution of vortices.  

Note that, the current profile will be more complicated for the more realistic case of a discrete vortex lattice, pinned locally disorder. However, by the above considerations, we expect the overall magnitude of the supercurrents at the edge to be roughly independent of these details.  In this light, it is difficult to understand how the presence or absence of vortices can account for the substantial saturation and hysteresis effects observed in \cite{Krupke}.

\begin{figure}[ttt]
\begin{center}
\includegraphics[width = 2.5in]{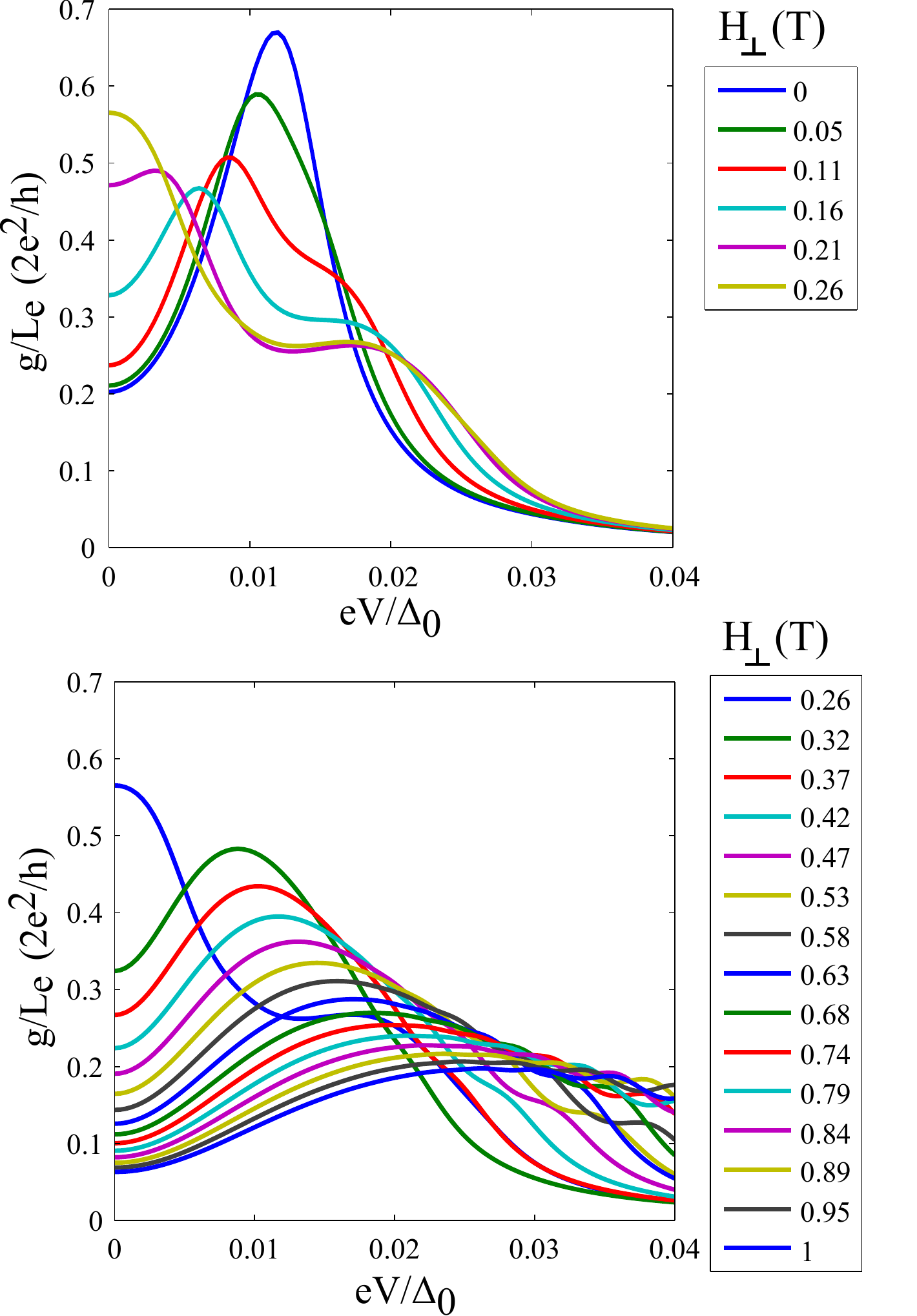}
\end{center}
\vspace{-.2in}
\caption{Tunneling peak splitting in a perpendicular magnetic field for $L_y = 900$, $t = 240$meV, $\Delta_0 = 10meV$, $U=2eV$, and $\gamma_0 = 0.4\Delta_0$.  Zeeman splitting is neglected and only positive bias is shown for simplicity.  At zero field, the peak is split by the spontaneous edge FM.  For $H<H_c^\text{edge}\approx 0.26 T$ (top panel), the zero field peak splits further due to the screening currents induced by the orbital field.  The spontaneous FM is killed in a field induced first order transition occurs at $H=H_c^\text{edge}$.  For $H>H_c^\text{edge}$ there is only a single split peak.
}
\vspace{-.2in}
\label{appfig:HperpTunnelingCurves}
\end{figure} 

\begin{figure}[ttt]
\begin{center}
\includegraphics[width = 2.5in]{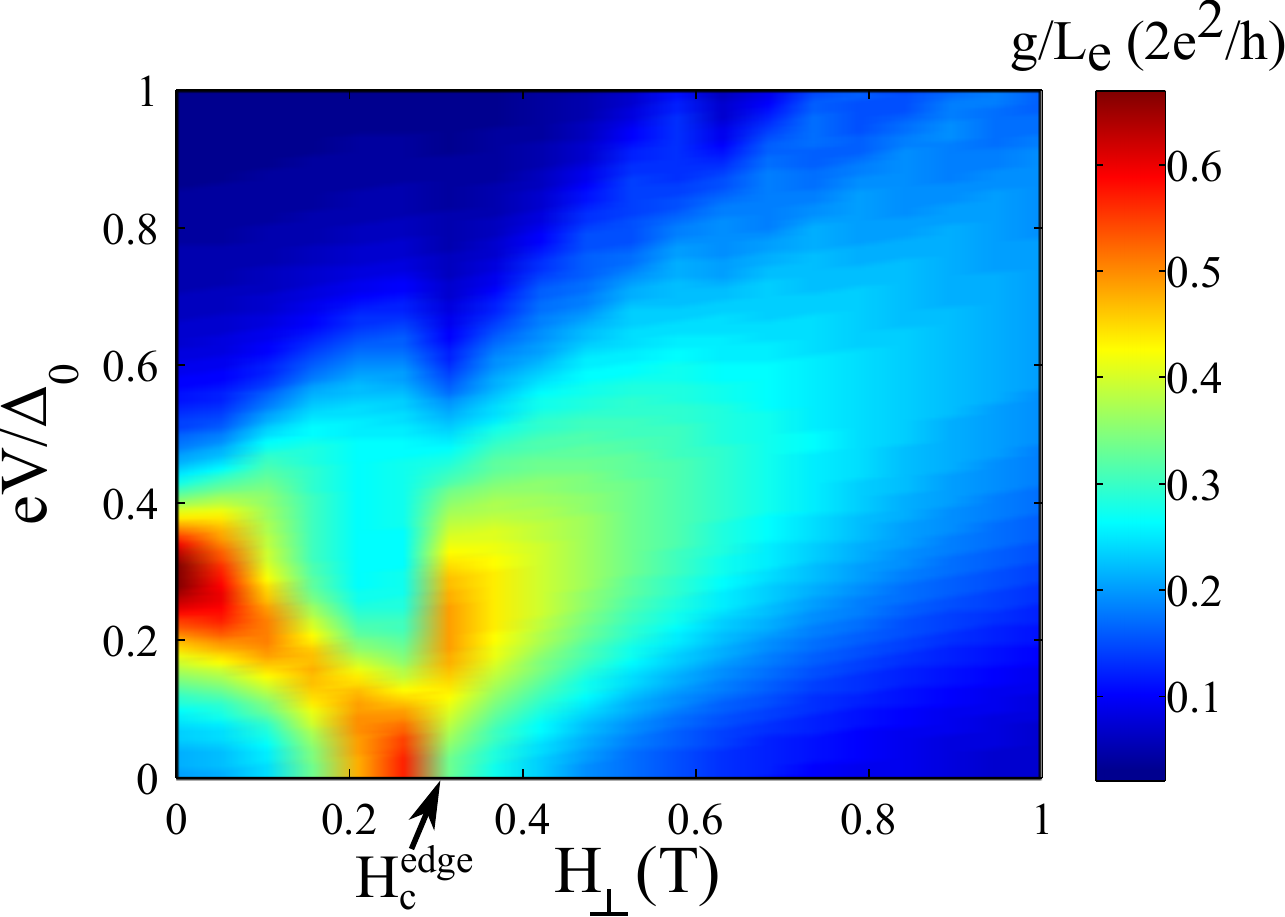}
\end{center}
\caption{Color plot of the field dependence of tunneling conductance, corresponding to Fig.~\ref{appfig:HperpTunnelingCurves}.
}
\vspace{-.2in}
\label{appfig:HperpTunnelingColor}
\end{figure}

\vspace{0.2in}\noindent\textbf{\small F.2 - Tunneling Conductance in Perpendicular Field}\\
Simulations of the tunneling conductance, due to Andreev reflection from the edge-states, in the presence of a perpendicular field are shown in Figures \ref{appfig:HperpTunnelingCurves} and \ref{appfig:HperpTunnelingColor}, for the case where there is spontaneous edge FM at zero-field. The top panel of Fig.~\ref{appfig:HperpTunnelingCurves} shows results for $H<H_{c}^\text{edge}\approx 260mT$, where the initially FM-split peak is further split into two by the supercurrents induced by the external field.  For larger fields, $H>H_c^\text{edge}$, the zero-field FM splitting is discontinuously destroyed by the applied field, resulting in a single split peak (see bottom panel of Fig.~\ref{appfig:HperpTunnelingCurves}).  At high-fields, the single peak continues to move out with increasing $H_\perp$, and broadens substantially.  This is in reasonable agreement with experiments, and the substantial broadening may obscure the visibility of the predicted low-field first order transition.

The corresponding tunneling data is shown in a false color plot in Fig.~\ref{appfig:HperpTunnelingColor}.


\begin{thebibliography}{10}
\bibitem{Endnote:Gwave}
It was also suggested\cite{Fogelstrom} that more exotic g-wave pairing with 8 nodes could be favored by spin-fluctuation mediated pairing produced by repulsion. However, we expect $d+ig$ edge-pairing due to be less favorable than the FM proposed here, since 1) the presence of extra nodes reduces the pairing energy gain, and 2) this high-angular momentum pairing is very susceptible to pair breaking from impurities and edge roughness.
\bibitem{Endnote:Hysteresis}
Ref.~\cite{Krupke} proposes an explanation of this hysteresis based on a surface barrier to vortices flowing in or out of the sample\cite{Bean}.
However, it is unclear to us how these barrier effects can explain the direction and other details of the hysteresis loop observed in \cite{Krupke}.  In particular, contrary to the arguments presented in \cite{Krupke}, the typical super-current near the edge is expected to increase linearly with the applied field: $j_\text{edge}\approx \frac{H_\perp}{\lambda_L}$, independently of whether or not there are vortices in the sample (see Appendix F.1).
\end{thebibliography}
\end{document}